\documentstyle[aps,preprint,epsfig]{revtex}

\begin{document}

\preprint{
\font\fortssbx=cmssbx10 scaled \magstep2
\hbox to \hsize{
\hbox{\fortssbx University of Wisconsin - Madison}
\hfill$\vcenter{\hbox{\bf MADPH-96-950}
                \hbox{\bf IFUSP 1225}
                \hbox{\bf hep-ph/9607324}
                \hbox{July 1996}}$ }}

\title{\vspace*{.5in}
 Prompt Charmonium Production in $Z$ Decays }

\author{
          E.\ M.\ Gregores\thanks{Email: gregores@phenos.physics.wisc.edu},
          F.\ Halzen\thanks{Email: halzen@phenxh.physics.wisc.edu}
        }

\address{Department of Physics, University of Wisconsin, Madison, WI
  53706}

\author{
        O.\ J.\ P.\ \'Eboli\thanks{ E-mail: eboli@fma.if.usp.br}
        }

\address{ Instituto de F\'{\i}sica, Universidade de S\~ao Paulo, C.P.\
66318, CEP 05389-970 S\~ao Paulo, Brazil.}

\maketitle

\thispagestyle{empty}

\begin{abstract}

The color-evaporation model quantitatively describes all data on
photoproduction and hadroproduction of charmonium.  Although the model
is in part nonperturbative, the associated parameters can for instance
be determined from the charmonium photoproduction data. At this point
its predictions for the prompt production of $\psi$'s at the $Z$ pole
are made with no free parameters.  We show here that this approach
successfully describes all data on the production of prompt $\psi$'s
in $Z$ decays.

\end{abstract}

\newpage

\section{Introduction}

There has been a renewed interest in studying the mechanism by which charmonium
is produced, triggered mostly by some puzzling data from the Fermilab Tevatron.
The standard perturbative QCD calculations using the color-singlet model failed
to explain the data, occasionally by orders of magnitude \cite{review}. The
data incited a complete review of the treatment of color in QCD and can, in
fact, be explained by allowing perturbative color octet $c\bar{c}$ states to
evolve into the asymptotic colorless charmonium states. This prescription is
present in both the color-evaporation\cite{cem,fh:1a,fh:1b,gor} and in the
color-octet models\cite{bbl}.

At the $Z$ resonance the majority of $\psi$'s are produced via $b$-hadron
decays, with a branching ratio B$(Z \rightarrow \psi + X)$ of approximately
$4.0\times10^{-3}$ \cite{bdecay}. Recently the OPAL collaboration also
presented evidence for the prompt production of $\psi$ in hadronic $Z$ decays
\cite{opal}.

We study here the prompt charmonium production in $Z$ decays in the framework
of the color-evaporation model. The predictive power of the model stems from
the fact that its free, nonperturbative parameters can be fixed by charmonium
photoproduction data, which leads to very well defined predictions for the
inclusive charmonium branching ratios. Although part of the mechanism is
essentially nonperturbative, it can be described by a minimal set of free
parameters which are few in comparison with the color-octet model.

We show that the color-evaporation model predicts an inclusive branching ratio
of $Z$ into prompt $\psi$ in agreement with the LEP results \cite{opal,lep}.
The dominant process is $Z \rightarrow c \bar{c} q\bar{q}$. It predicts an
energy spectrum which is very different from the
one predicted by the color-singlet model as well as a substantially larger
branching ratio of $Z$ into charmonium.


\section{The Model and Its Parameters}

The color-evaporation approach, which actually predates the color-singlet
approach, quantitatively describes all charmonium photo- and hadroproduction
data \cite{aegh1,aegh2}. The model simply states that charmonium production is
described by the same dynamics as $D \bar{D}$ production, {\em i.e.}, by the
formation of a colored $c\bar{c}$ pair.  Rather than imposing that the
$c\bar{c}$ pair is in a color-singlet state in the short-distance perturbative
diagrams, it is argued that the appearance of color-singlet asymptotic states
solely depends on the outcome of large-distance fluctuations of quarks and
gluons. These large-distance fluctuations are probably complex enough for the
occupation of different color states to approximately respect statistical
counting. In other words, the formation of color-singlet states is a
nonperturbative phenomenon. In fact, it does not seem logical to enforce the
color-singlet property of the $c \bar{c}$ pair at short
distances, given that there is an infinite time for soft gluons to readjust the
color of the pair before it appears as an asymptotic $\psi$, $\chi_c$ or,
alternatively, $D \bar{D}$ state.  It is indeed hard to imagine that a
color-singlet state formed at a range $m_{\psi}^{-1}$ automatically survives to
form a $\psi$.  This formalism was proposed almost twenty years ago
\cite{cem,fh:1a,fh:1b,gor} and subsequently abandoned for no good reason.

In the color-evaporation model the sum of the cross sections of all
onium and open charm states is described by
\begin{equation}
\sigma_{\rm onium} = \frac{1}{9} \int_{2 m_c}^{2 m_D} dm~
\frac{d \sigma_{c \bar{c}}}{dm} \; ,
\label{sig:on}
\end{equation}
and
\begin{eqnarray}
\sigma_{\rm open} &=& \frac{8}{9}  \int_{2 m_c}^{2 m_D} dm~
\frac{d \sigma_{c \bar{c}}}{dm}
+ \int_{2 m_D} dm~\frac{d \sigma_{c \bar{c}}}{dm} \; ,
\label{sig:op}
\end{eqnarray}
where the cross section for producing heavy quarks, $\sigma_{c
  \bar{c}}$, is computed perturbatively, irrespective of the color of
the $c \bar{c}$ pair, and $m$ is the invariant mass of the $c \bar{c}$
pair.  The coefficients $\frac{1}{9}$ and $\frac{8}{9}$ represent the
statistical probabilities that the $3\times\bar3$ charm pair is
asymptotically in a singlet or octet state \cite{aegh1}.

The color-evaporation model assumes a factorization of the production of the
$c\bar{c}$ pair, which is perturbative and process dependent, and the
materialization of this pair into a charmonium state by a mechanism that is
nonperturbative and process independent.  This
assumption is reasonable given that the characteristic time scales of the two
processes are very different: the time scale for the production of the pair is
the inverse of the heavy-quark mass, while the formation of the bound state is
associated to the time scale $1/\Lambda_{\rm
QCD}$.  Comparison with the $\psi$ data requires knowledge of the fraction
$\rho_\psi$ of produced onium states that materialize
as $\psi$'s, {\em i.e.,}
\begin{equation}
\sigma_\psi = \rho_\psi \sigma_{\rm onium} \; ,
\label{frac}
\end{equation}
where $\rho_\psi$ is assumed to be a constant.  This assumption is in agreement
with the low-energy data \cite{gksssv,schuler}. Notice that a single
nonperturbative factor $\rho$ describes a given charmonium state, regardless of
the spin and orbital angular momentum of the charm pair. Therefore the
color-evaporation model is more economical in nonperturbative parameters than
the color-octet mechanism.

In Ref.~\cite{aegh2}, we determined the factor $\rho_\psi$ from an analysis of
photoproduction of charmonium. Using this value we were able to accommodate all
data on the hadroproduction of charmonium without introducing any new
parameters. This is a very remarkable result given that the subprocess
responsible for the charmonium hadroproduction changes from $q \bar{q}$ fusion
to $g g$ fusion as the center-of-mass energy is increased. The fragmentation
factor $\rho_\psi$ turned out to be 0.50 (0.43) when we adopted the GRV 94 HO
(MRS-A) parton distribution functions for the proton \cite{grv}
(\cite{mrsa}).


\section{Prompt decay of $Z$ into charmonium}

In the color-evaporation model the width for inclusive $Z$ decay into prompt
charmonium is:
\begin{equation}
\Gamma( Z \rightarrow \mbox{prompt charmonium}) =
\frac{1}{9} \int_{2 m_c}^{2 m_D} dm~
\frac{d \hat{\Gamma}_{c \bar{c}}}{dm} \; ,
\end{equation}
where $\hat{\Gamma}$ is the partonic width for producing a $c \bar{c}$
pair. In order to obtain the partial width into a specific charmonium
state we multiply the above expression by the appropriate
fragmentation factor $\rho$, which was determined from charmonium
photoproduction data \cite{aegh1,aegh2}.  Notice that the predictions
for the $Z$ decay into charmonium are parameter-free.  Therefore the
production of charmonium at the $Z$ pole is a clean test of the
validity of the color-evaporation model.

We have evaluated all the tree-level partonic amplitudes using the package
MADGRAPH \cite{tim}. The leading-order process in $\alpha_s$ is $Z \rightarrow
c\bar{c} g$, which leads to the production of a charmonium state and a hard
jet.  The partial width $Z \rightarrow \psi g$ is only $6 \times 10^{-7}$ GeV,
for $m_c=1.45$ GeV, $\rho_\psi = 0.50$, and $\alpha_s(2 m_c) = 0.235$. It is
small because the virtual quark propagator suppresses the amplitude by a factor
of the order $m_c/m_Z$. This situation persists even when taking into account
the next-to-leading order corrections.  Therefore, this decay mode is too small
to be seen at
LEP I.

\begin{figure}
\vglue.25in
\centering
\mbox{\epsfig{file=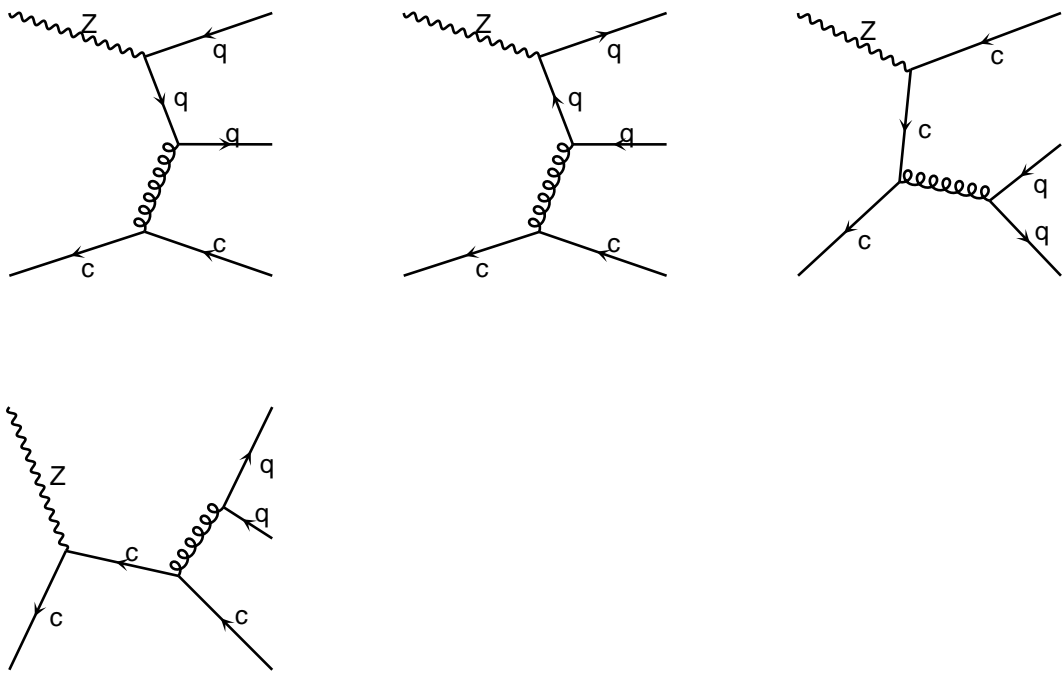,width=.85\linewidth}}

\bigskip
\caption{Feynman diagrams leading to $Z \rightarrow c \bar{c} q \bar{q}$.
In the  case $q=c$ we must also add the crossed diagrams.}
\label{zccqq}
\end{figure}

Although formally of higher order in $\alpha_s$, the dominant process for the
inclusive decay of the $Z$ into charmonium is $Z \rightarrow c \bar{c} q
\bar{q}$, where $q=u$, $d$,
$s$, $c$, and $b$. We show in Fig.\ \ref{zccqq} the Feynman diagrams
contributing to this process. Table~I contains the color-evaporation-model
predictions for the $Z$ partial width into the different partonic final states
\cite{note}. In order to access the uncertainties in the
determination of the fragmentation factor $\rho_\psi$ from the photo-production
of $\psi$ data, we considered the values $\rho_\psi = 0.50$ and $0.43$ which
were obtained from $\psi$ photoproduction using the GRV 94 HO structure
functions with $m_c=1.45$ GeV and MRS-A distribution functions with $m_c=1.43$
GeV, respectively \cite{aegh2}. Notice that the branching fraction of $Z$ into
prompt $\psi$ is $(1.7\mbox{--}1.8)\times10^{-4}$. This is to be contrasted
with the color-singlet model which predicts a branching fraction for
direct $\psi$ in $Z$ decay of the order $3\times10^{-5}$ \cite{bcy}. The
color-evaporation model leads to a branching fraction larger by almost an order
of magnitude.

Using vertex detectors the LEP collaborations have recently been able to
separate the prompt production of $\psi$ from those originating from $b$-hadron
decays\cite{opal,lep}. The OPAL collaboration reports that
\[
\text{B}( Z \rightarrow \text{prompt}~ \psi + X) =
(1.9 \pm 0.7 \pm 0.5 \pm 0.5) \times 10^{-4} \; .
\]
Therefore, the color-evaporation-model prediction is in excellent agreement
with the LEP results, illustrating how this approach gives a complete picture
of the charmonium production in hadron-hadron, $\gamma$-hadron, and $Z$ decays.
At this point we should point out that the color-octet model predicts a
branching ratio $Z \rightarrow \psi + X$ to be $2.9\times10^{-4}$
\cite{opal}, which is also in agreement with the experimental data.

\begin{figure}
\centering
\mbox{\epsfig{file=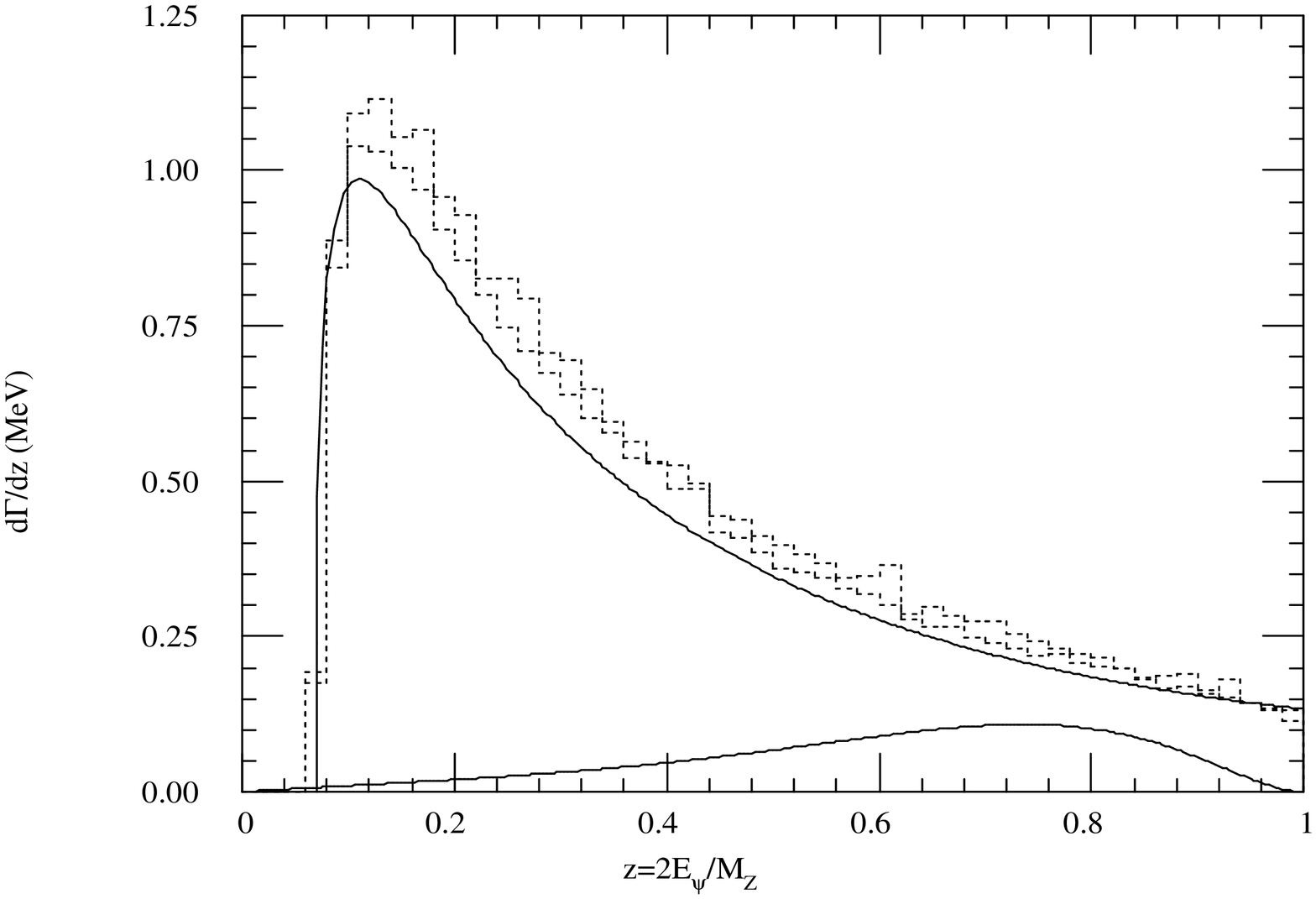,width=.85\linewidth,angle=-90}}

\bigskip
\caption{Energy spectrum $d\Gamma/dz$ of the $\psi$ from
  the dominant partonic process $Z \rightarrow c \bar{c} q \bar{q}$.
  The upper (lower) solid curve stands for the prediction of the color
  octet (singlet) model according to Ref.\ \protect\cite{cky}.  The
  upper (lower) dashed curves is the distribution predicted by the
  color evaporation approach when we take $\rho_\psi=0.50$ (0.43),
  $m_c=1.45$ (1.43) GeV, and $\alpha_s(2 m_c) = 0.235$ (0.236).
}
\label{dgdz}
\end{figure}

In Fig.\ \ref{dgdz} we show the energy distribution ($d\Gamma/dz$ with
$z=2 E_\psi/M_Z$) of prompt $\psi$ for the dominant process
$Z\rightarrow c\bar{c} q\bar{q}$, where we summed over all quark
flavors.  The band between the dashed curves is the distribution
predicted by the color-evaporation approach, when we consider the
different values for $\rho_\psi$ given above in order to access the
theoretical uncertainties. For comparison, we show the same
distribution for the dominant process ($Z \rightarrow \psi q\bar{q}$)
computed in the context of the color-octet (singlet) model represented
by the upper (lower) solid curve \cite{cky}.

As we can see from Fig.\ \ref{dgdz}, the color-evaporation
distribution is softer than the one predicted by the color-singlet
model; it dominates however over the entire allowed range
for~$z$. Moreover, the color-evaporation mechanism and the color-octet
model predict very similar distributions. Taking into account the
theoretical uncertainties in the determination of the nonperturbative
parameters describing these models, it is impossible to distinguish
between them on the basis of this distribution. Therefore the
measurement of the energy spectrum of the prompt $\psi$ is not a
distinctive signature of the color-octet model, contrary to what is
claimed in Ref.\ \cite{cky}. In the OPAL analysis
of the production of prompt $\psi$'s~\cite{opal}, it is shown that the
color-octet model describes the momentum distribution of
prompt~$\psi$. Since this model and the color-evaporation mechanism have
similar predictions for this distribution, we can conclude that this
last model also describes the available data.


\section{Conclusions}

The predictions of the color-evaporation model for the production of
prompt $\psi$ on $Z$ decays have no free parameters once we use the
photoproduction of charmonium to constrain the nonperturbative
parameters of the model. We showed that this approach gives a good
description of the available data on the production of prompt $\psi$
at LEP I. Taking into account the success of this model to describe
the photo- and hadroproduction of charmonium, we can conclude that
this model gives a robust and simple parametrization of all charmonium~physics.

We would like to stress that the color-evaporation model has the same
degree of success as the color-octet mechanism; however, the number of
free parameters in the color-evaporation approach is smaller. The
$\psi$'s produced through the color-evaporation mechanism are
basically unpolarized since the polarization information is lost
because of the multiple soft-gluon exchanges \cite{mirkes}. On the
other hand, the (non)polarization of $\psi$ is hard to explain in the
framework of the color-octet model
\cite{pol:com,braatenchen}. Therefore, the measurement of the
polarization of the produced charmonium may very well be a tool to
discriminate between these competing descriptions.


\acknowledgments

This research was supported in part by the University of Wisconsin
Research Committee with funds granted by the Wisconsin Alumni Research
Foundation, by the U.S.\ Department of Energy under grant
DE-FG02-95ER40896, by Conselho Nacional de Desenvolvimento
Cient\'{\i}fico e Tecnol\'ogico (CNPq), and by Funda\c{c}\~ao de
Amparo \`a Pesquisa do Estado de S\~ao Paulo (FAPESP).



\begin{table}[htb]
\label{t1}
\caption{Partial decay widths for $Z \rightarrow \psi ~+~ X$.
  The results were obtained using $\rho_\psi=0.50$ [0.43], $m_c=1.45$
  [1.43] GeV, and $\alpha_s(2 m_c) = 0.235$ [0.236].}
\begin{center}
\begin{tabular}{|c|c|c|}
$X$ &  $\Gamma( Z \rightarrow \psi ~+~ X)$ (GeV)
& Br$(( Z \rightarrow \psi ~+~ X)$
\\ \hline
$g$ & 6.0~10$^{-7}$ [5.6~10$^{-7}$]
& 2.4~10$^{-7}$ [2.3~10$^{-7}$]
\\ \hline
$u \bar{u}$, $d \bar{d}$, $s \bar{s}$ & 2.8~10$^{-4}$ [2.6~10$^{-4}$]
& 1.1~10$^{-4}$ [1.0~10$^{-4}$]
\\ \hline
$c \bar{c}$  & 7.8~10$^{-5}$ [7.3~10$^{-5}$]
& 3.1~10$^{-5}$ [2.9~10$^{-5}$]
\\ \hline
$b \bar{b}$  & 8.4~10$^{-5}$ [7.9~10$^{-5}$]
& 3.4~10$^{-5}$ [3.2~10$^{-5}$]
\\ \hline
$q \bar{q}$ (all $q$) & 4.4~10$^{-4}$ [4.1~10$^{-4}$]
& 1.8~10$^{-4}$ [1.7~10$^{-4}$]
\end{tabular}
\end{center}
\end{table}

\newpage

\begin{references}
\bibitem{note}The partonic process $Z \rightarrow c\bar{c} c\bar{c}$
  also leads to the production of two $\psi$'s. However, the partial
  width into this final state is only $1.3 \times 10^{-9}$ GeV.
\end{references}
\end{document}